\documentclass[aps,pra,twocolumn]{revtex4}
\usepackage{amsmath}
\usepackage{amssymb}

\newcommand\beq{\begin{equation}}
\newcommand\eeq{\end{equation}}
\newcommand\bea{\begin{eqnarray}}
\newcommand\eea{\end{eqnarray}}
\newcommand\nn{\nonumber}

\newcommand\ro{\hat\rho}
\newcommand\Po{\hat P}
\newcommand\Qo{\hat Q}
\newcommand\Ho{\hat H}
\newcommand\Lo{\hat L}
\newcommand{\Ro}{\widehat\rho}
\newcommand\dg{^\dagger}

\newcommand\prr{^{\prime\prime}}

\newcommand\ketx{\vert x\rangle}
\newcommand\brax{\langle x\vert}

\newcommand\bray{\langle y\vert}
\newcommand\eps{\epsilon}
\newcommand\der{\partial}

\newcommand\tr{\mathsf{tr}}
\newcommand\Herm{\mathbb{H}}
\newcommand\HC{\mathsf{h.c.}}
\newcommand\Real{\mathbb{Re}}
\newcommand\Ima{\mathbb{I}}
\newcommand\Mean{\mathbb{M}}
\newcommand\rank{\mathsf{rank}}
\newcommand\range{\mathsf{range}}

\newcommand{\al}{\alpha}
\newcommand{\be}{\beta}
\newcommand{\si}{\sigma}
\newcommand{\la}{\lambda}

\newcommand{\Loa}{\Lo_\al}

\newcommand{\Laxy}{\Loa(x,y)}
\newcommand{\Laxyd}{\Lo\dg_\al(x,y)}
\newcommand{\Layx}{\Loa(y,x)}
\newcommand{\Layxd}{\Lo\dg_\al(y,x)}
\newcommand\DQ{D_\mathrm{Q}}
\newcommand\DC{D_\mathrm{C}}
\newcommand\GCQ{G_\mathrm{CQ}}
\newcommand\GCQb{\overline{G}_\mathrm{CQ}}
\newcommand\rCQ{r_\mathrm{CQ}}
\newcommand{\bD}{\mathcal{D}}
\newcommand\FQC{F_\mathrm{QC}}
\newcommand\HCl{H_\mathrm{C}}
\newcommand\HoQ{\Ho_\mathrm{Q}}
\newcommand\HoCQ{\Ho_\mathrm{CQ}}
\newcommand{\rox}{\ro_x}

\newcommand{\Taxy}{T_\al(x,y)}
\newcommand{\Tayx}{T_\al(y,x)}
\newcommand\Hfr{\Ho_\mathrm{fr}}
\newcommand\sio{\hat\si}
\newcommand\sv{\mathbf{s}}
\newcommand\so{\hat s}
\newcommand{\sqrtD}{\sqrt{g}}

\begin{document}
\title{Hybrid completely positive Markovian quantum-classical dynamics}    
\author{Lajos Di\'osi}
\affiliation{Wigner Research Center for Physics, H-1525 Budapest 114 , P.O.Box 49, Hungary}
\affiliation{E\"otv\"os Lor\'and University, H-1117 Budapest, P\'azm\'any P\'eter stny. 1/A}
\date{\today}

\begin{abstract}
A concise and self-contained derivation of hybrid quantum-classical dynamics
is given in terms of Markovian master equations. Many previously known results
are re-derived, revised, some of them completed or corrected. Using as simple
method as possible, our goal is a brief introduction to state-of-the-art of 
hybrid dynamics, with a limited discussion of the implications for foundations, 
and without discussion of further relevance in measurement problem, quantum-gravity, or chemistry, numeric methods, etc.
Hybrid dynamics is defined as special case of composite quantum
dynamics where the observables of one of the two subsystems are restricted
for the commuting set of diagonal operators in a fixed basis. With this restriction,
the derivation of hybrid dynamical equations is clear conceptually and 
simple technically. Jump and diffusive dynamics follow in the form of hybrid 
master equations. Their stochastic interpretation (called unravellings) is derived.
We discuss gauge-type ambiguities, problems of uniqueness, and covariance of 
the diffusive master equation. Also conditions of minimum noise and of
monitoring the quantum trajectory are derived. We conclude
that hybrid formalism is equivalent with standard Markovian theory
of time-continuous quantum measurement (monitoring) on one hand, 
and is a motivating alternative formalism on the other hand. 

\end{abstract}
\maketitle

\section{Introduction}
In real word,  quantum and classical phenomena coexist and evolve according
to their own rules. They do interact, of course, and we know very well the 
mathematical models of  some particular interactions. The action 
of a classical system on the quantum one is modeled if we make the Hamiltonian
$\Ho$ depend on the classical system's variables $x$. The \emph{backaction},
i.e. the quantum system's impact on a classical one is also known well
from the quantum measurement: the quantum system rules the pointer 
position $x$ of a classical meter. This backaction is extremely specific.
More general backaction is the central problem to understand and to describe
mathematically when quantum and classical systems are interacting. 

The central object of composite quantum-classical systems is the hybrid state,
represented by the hybrid density \cite{aleksandrov1981}:
\beq\label{rox}
\ro(x)=\ro_x\rho(x),
\eeq 
where $\rho(x)$ is the normalized probability density of the classical
variables $x$ and $\ro_x$ is the density operator of the quantum system
conditioned on the value $x$ of the classical variable. 

The efforts and
results concern the possible evolution equations for $d\ro(x)/dt$. 
The bottleneck is the backaction although its elementary
pattern is known from all quantum theory textbooks.
The von Neumann measurement of the complete orthogonal
set $\{\Po_x\}$ of Hermitian projectors is imposes the following
change of the hybrid state \eqref{rox}:
\begin{widetext}
\beq\label{backaction}
\Bigl\{\ro\rightarrow\frac{\Po_x\ro\Po_x}{\tr(\Po_x\ro)}
\mbox{ with prob }\rho(x)=\tr(\Po_x\ro)\Bigr\}
\Longleftrightarrow
\Bigl\{\ro(x)\longrightarrow\Po_x\rox\Po_x\Bigr\},
\eeq 
\end{widetext}
As we see, the textbook stochastic jumps are \emph{equivalent} 
with a single deterministic map of the hybrid state. If we construct a stochastic dynamics underlying the process
on the lhs, we have an equivalent determinsitic hybrid dynamics for $\ro(x)$.
Von Neumann's statistical interpretation of quantum states (also called
the Born rule) follows from the statistical interpretation of the hybrid state.

For longtime, the dynamics of von Neumann measurement,
l.h.s. of \eqref{backaction}, was missing from
the textbooks, as irrelevant. Lately, by very different motivations, it was
constructed in the continuous limit of discrete von Neumann measurements.
The wining formalism has been Markovian stochastic equations, not 
the hybrid formalism. Refs. \cite{gisin1984quantum,diosi1988ito,belavkin1988}
were milestones, reviews are in \cite{wiseman2001diosi,brun2002simple,breuer2002theory,jacobs2006,wiseman2009quantum}.

Whether hybrid dynamics, developped on their own, yield more than
the time-continuous measurements in hybrid formalism, 
i.e.: time-continuous extension of the r.h.s. of  \eqref{backaction}? 
Before the answer,  we introduce ourselves into
the calculus of hybrid dynamics.

A large body of investigations of quite different concepts, motivations, methods, level
of mathematical rigor, etc., emerged in forty years, from the earliest 
attempts to couple quantum and canonical classical systems 
\cite{aleksandrov1981,gerasimenko1982,boucher1988}
through the present author's 
\cite{diosi1995twoplanck,diosi1996comment,diosi1998,diosi1999hybrid,diosi2000,diosi2011,
diosi2012freewill,diosi2014hybrid}
and other's results
\cite{blanchard1993,blanchard1995,alicki2003,
oppenheim2018post,oppenheim2022constraints,oppenheim2022two,
oppenheim2022gravitationally,
layton2022healthier,oppenheim2023objective}
directly related to the present work, and many other contributions, e.g., in refs.  
\cite{anderson1995,prezhdo1997mixing,peres2001hybrid,salcedo2012,tilloy2018binding,
kapral2006, barcelo2012hybrid,elze2012linear,
elze2013quantum,buric2014phase,hall2016consistency,
bondar2019,donadi2022seven}.

Here we use a unique approach, elementary derivations, economic
presentation,  in order to give a complete but concise account of
Markovian hybrid quantum-classical dynamics modeled by hybrid master equations (HMEs).
Sec. \ref{sec_HME} derives the canonical jump HME and its 
stochastic interpretation, called \emph{unraveling}.
From this HME, sec. \ref{sec_diffHME} derives the limit of continuous,
diffusive HME which, in sec. \ref{sec_covHME} is put into the general
covariant form. Condition of minimum irreversibility, covariant and
non-covariant unravelings, and monitoring  are contained in the subsections. Comparison with and comments on previous results will be concentrated 
in the sec. \ref{sec_disc}. 
At last, sec. \ref{sec_concl} contains conclusion of the author. 
Appendix \ref{app1} shows an elemantary example, \ref{app2}
contains application to canonical classical subsystems, with a simplest example. %

\section{Hybrid master equation}\label{sec_HME}
Quantum theory is universal in our approach and classical systems are special
case of quantum ones: operators are restricted for a commutative
subset, i.e., they are diagonal in a fixed basis. 
Accordingly, consider the Hilbert space
$\mathcal{H}_{QC}=\mathcal{H}_Q\otimes\mathcal{H}_C$
where $\mathcal{H}_C$ will host our classical system in a fixed basis 
$\{\ketx\}$. The composite quantum state $\Ro$, and composite Hamiltonian
$\widehat{H}$ (as well as the composite observables) are diagonal
in the fixed basis:
\bea\label{hybs}
\Ro&=&\sum_x\ro(x)\otimes\ketx\brax,\\
\widehat{H}&=&\sum_x\Ho(x)\otimes\ketx\brax.
\eea
The  block-diagonality of the objects $\ro(x),\Ho(x)$ ensures classicality
of the subsystem in $\mathcal{H}_C$, i.e., that its state is and remains diagonal
in the fixed basis. %
The block-diagonal objects $\ro(x),\Ho(x)$ will be called respectively the
hybrid state and hybrid Hamiltonian. 
Note the notational difference: operators on $\mathcal{H}_{QC}$ wear 
wider hats than operators on $\mathcal{H}_{Q}$. %
We are looking for a Markovian evolution equation
for $\Ro$, which is completely positive (CP) and preserves the
block-diagonal form of $\Ro$. 

Since CP maps represent the most general quantum dynamics, %
we  start from the CP map $\Lambda$ of the composite state $\Ro$
and request that it preserve block-diagonality, i.e., classicality of
the subsystem in $\mathcal{H}_C$. Conveniently,  we
can write the general form of the map for the hybrid representation $\ro(x)$
of $\Ro$.
With Einstein convention of summation, it reads
\beq\label{Kraushyb}
\ro_\Lambda(x)=\sum_y D_{\be\al}(x,y)\Loa\ro(y)\Lo\dg_\be,
\eeq
where $\{\Loa\}$ is an operator basis in $\mathcal{H}_Q$. 
The Hermitian matrix $D$ satisfies
\beq
\sum_x D_{\be\al}(x,y)\Lo\dg_\be\Loa=\hat{I}
\eeq
for all $y$.
If we diagonalize $D$ at each point $(x,y)$, the form of the map becomes
\beq\label{Kraushybd}
\ro_\Lambda(x)=\sum_{y,\al} \la_\al(x,y)\Loa(x,y)\ro(y)\Loa\dg(x,y),
\eeq
where $\{\Loa(x,y)\}$ is an operator basis depending on $(x,y)$. Hence 
all $\la_\al(x,y)\geq0$ for all $\al$ and $(x,y)$ otherwise the map 
$\Lambda$ cannot be CP. Now we can absorb each factor $\la_\al$ 
into $\Lo_ \al$ and return to the full operator formalism. We can write 
the CP map, preserving the block-diagonal form of $\Ro$ into the standard
quantum mechanical form
\beq\label{Kraus}
\Ro_\Lambda=\widehat{L}_\al\Ro\widehat{L}_\al\dg,
\eeq
with the linearly independent Kraus-operators defined by
\beq
 \widehat{L}_\al=\sum_y\Laxy\otimes\ketx\bray,
\eeq
where 
$\Lo_1(x,y),\Lo_2(x,y),\dots$ are linearly independent.
If $\Lambda$ is a semigroup and generates time evolution of $\Ro$
then, according to the GKLS theorem \cite{lindblad1976,gorini1976}, the  evolution is governed
by the quantum master equation of the following from:
\beq\label{ME}
\frac{d\Ro}{dt}=-i[\widehat{H},\Ro]+\widehat{L}_\al\Ro\widehat{L}_\al\dg
                                                 -\Herm \widehat{L}_\al\dg\widehat{L}_\al\Ro,
\eeq
where the $\widehat{L}_\al$'s are called the Lindblad generators.
They can (and will) be chosen such that $\widehat{I},\widehat{L}_1,\widehat{L}_2,\dots$ are linearly independent. %
As expected, this quantum master equation 
preserves block-diagonality by construction hence it obtains a closed
equation in the hybrid formalism
\cite{blanchard1993,blanchard1995,alicki2003,diosi2014hybrid,oppenheim2022two,
layton2022healthier,oppenheim2023objective}:
\begin{widetext}
\beq\label{HME}
\frac{d\ro(x)}{dt}=-i[\Ho(x),\ro(x)]
+\sum_y\left(\Laxy\ro(y)\Laxyd-\Herm\Layxd\Layx\ro(x)\right),
\eeq
\end{widetext}
where $\Herm$ means the Hermitian part of the subsequent expression.
This is the canonical form of the CP Markovian HME where the classical
system is discrete.
The hybrid generators are only restricted by their linear independence
and their linear independence of $\hat{I}\delta(x,y)$ as well. 
Their dependence on $(x,y)$ can otherwise be arbitrary. %
Through our work $\delta(x,y)$ will denote the usual discrete delta-function. %
The backaction is encapsulated by the generators $\Laxy$.
Note the basic lesson: when the quantum-classical interaction is mutual
influence between the subsystems then the hybrid system can never 
be reversible, the evolution is governed by hybrid master (kinetic)
equations. Quantitative lower bounds on their irreversibility (noise)
will be derived in sec. \ref{sec_diffHME}.

The choice of the hybrid Hamiltonian and the hybrid generators is
unique upto arbitrary complex functions $\ell_\al(x)$ since 
the HME \eqref{HME} is invariant for the following shifts:
\bea\label{shift}
\Laxy&\rightarrow&\Laxy+\ell_\al(x)\delta(x,y),\nn\\
\Ho(x)&\rightarrow&\Ho(x)-\frac{i}{2}\bigl(\ell_\al^\star(x)\Loa(x,x)-\HC\bigr),
\eea
Special cases may allow much larger groups of such `gauge' freedom when
the  dependence of the generators on $(x,y)$ is degenerate, 
like in secs. \ref{sec_diffHME}, \ref{sec_covHME} of continuous HMEs.

\subsection{Stochastic unravelling}\label{sec_unr}
Master equations are deterministic.
Like any quantum or classical master equation, also our HME \eqref{HME} possesses 
statistical interpretation. The solution $\rho(x)$ of a classical master (kinetic)
equation can be decomposed into a unique stochastic process $x_t$ of random trajectories.
Quantum master equations can be decomposed into stochastic 
processes $\psi_t$ of random quantum trajectories, the decomposition is called unraveling. 
Quantum unravellings are not unique. 
Here we define the hybrid of the classical and quantum 
stochastic decompositions (unravellings).

Let us formulate the mathematical condition of hybrid unraveling.
If $\ro(z,t)$ is the solution of the HME \eqref{HME} then it is the stochastic mean $\Mean$
over the contribution of hybrid trajectories $(x_t,\psi_t)$:
\beq\label{unrcond}
\ro(z,t)=\Mean\psi_t\psi_t\dg\delta(z,x_t),
\eeq
where the $x_t$ and $\psi_t$ are correlated jump stochastic 
processes, 
one in the space of classical cooordinates and the other in the 
Hilbert space of state vectors. %
Consider first the unique unraveling of the distribution  $\rho(x)=\tr\ro(x)$
of the classical subsystem. The trace of the HME \eqref{HME} yields the following
classical master (kinetic) equation:
\beq\label{kinetic}
\frac{d\rho(x)}{dt}=\sum_\al\sum_y\Bigl(\Taxy\rho(y)-\Tayx\rho(x)\Bigr),
\eeq
with the $\psi$-dependent transition (jump) rate from $x$ to $y$ for each $\al$:
\beq\label{rate}
\Tayx=\tr\left(\Layx\rox\Layxd\right)=\bigl\langle\Layxd\Layx\bigr\rangle_x.
\eeq
We introduce the total transition rate from $x$:
\beq\label{ratetot}
T(x)=\sum_\al\sum_y\Tayx.
\eeq
To unravel the quantum subsystem, we need the
anti-Hermitian (frictional) hybrid Hamiltonian defined by
\beq\label{Hfr}
-i\Hfr (x)=-\frac12\sum_y\Layxd\Layx.
\eeq
The corresponding frictional Schr\"odinger equation is
\beq\label{Schfr}
\frac{d\psi}{dt}=\left(-i\Ho(x)-i\Hfr (x)+\frac12 T(x)\right)\psi,
\eeq
where $\tfrac12 T(x)$ restores the norm $\psi\dg\psi$ 
since $\tfrac12 T(x)=-\Ima\langle\Hfr \rangle$,  
cf. eqs. (\ref{rate}-\ref{Hfr}); symbol $\Ima$ means the imaginary part.  

The hybrid unraveling consists of the following two correlated jump stochastic 
(piecewise deterministic) processes, one for $x_t$, another for $\psi_t$
(cf. \cite{oppenheim2023objective}):
\bea
x&=&\mbox{ const.},\label{unrdx}\\
\frac{d\psi}{dt}&=&\left(-i\Ho(x)-i\Hfr (x)\psi +\tfrac12 T(x)\right)\psi,\label{unrdpsi} 
\eea
\vskip-20pt
\beq\label{unrjumps}
\hskip-32pt\mbox{jumps}
\left\{\begin{array}{ccc}
x       &\rightarrow& x'~,\\
\psi &\rightarrow&\frac{\Loa(x,x')\psi}{\sqrt{T_\al(x',x)}}
\end{array}\right.
\mbox{at rate }T_\al(x',x).
\eeq
The proof that the unraveling satisfies the condition \eqref{unrcond}
is the following. With the notation $\Po=\psi\psi\dg$, the condition reads
\beq\label{unrcondP}
d\ro(z,t)=d \Mean\Po_t\delta(z,x_t),
\eeq
where $d\ro(z,t)$ is determined by the HME \eqref{HME}.
The  unraveling yields two terms for the change of $\Po\delta(z,x)$ in time $dt$:
\begin{widetext}
\begin{eqnarray}
\Po\delta(z,x)\rightarrow&&
\bigl(1-T(x)dt\bigr)
\times\left(\Po-i[\Ho(x),\Po]dt-i[\Hfr (x),\Po]_{+}dt+T(x)\Po dt\right)\delta(z,x)
\nn\\
   &&+\sum_\al T_\al(x',x)dt\times\frac{\Loa(x,x')\Po\Loa\dg(x,x')}{T_\al(x',x)}\delta(z,x').
\end{eqnarray}
\end{widetext}
The first term comes from the deterministic eqs. (\ref{unrdx},\ref{unrdpsi}),
the second term represents the average of jumps \eqref{unrjumps}.
The $\psi$-dependent transition rates $T_\al(x',x)$ and $T(x)$ cancel,
we are are left with an expression linear in $\Po\delta(z,x)$. 
Taking its mean obtains this:
\begin{widetext}
\beq
\frac{d}{dt}\Mean\Po\delta(z,x)=
\Mean\left(-i[\Ho(x),\Po]-i[\Hfr (x),\Po]_{+} \right)\delta(z,x)
+\Mean\Loa(x,x')\Po\Loa\dg(x,x')\delta(z,x')
\eeq
\end{widetext}
If we recall the expression \eqref{Hfr} of $\Hfr $ then we can recognize that
$\Mean\Po\delta(z,x)$ satisfies the HME \eqref{HME}.

The unraveling (\ref{unrdx}-\ref{unrjumps}) is ambiguous. The shifts \eqref{shift}
leave the HME \eqref{HME} invariant but we get  different unravelings.
Nevertheless, the unraveling becomes invariant and unique with the  
replacement (cf. \cite{diosi1986}): 
\beq\label{dio86}
\Laxy\rightarrow\Laxy-\langle\Laxy\rangle\delta(x,y).
\eeq
One would think that in eq. \eqref{unrjumps} the state $\psi_t$ and $x_t$ 
are jumping together. This is not true if $\Loa(x,x)\neq0$ since $\Loa(x,x)$
generates nontrivial jump of $\psi_t$ and no jump for $x_t$.
So, to synchronize the jumps of $\psi$ and $x$ either we consider
the invariant modification \eqref{dio86} of the unraveling or we just
request that $\Loa(x,x)$ vanishes for all $\al$ and $x$.

\subsection{Monitoring the quantum trajectory}\label{sec_mon}
The quantum trajectory $\psi_t$ is not observable in general since its
detection inevitably perturbs it.  That is the problem of \emph{monitoring} the
quantum system. Can we, just by monitoring the classical
subsystem's  $x_t$, monitor the evolution of $\psi_t$? Generally we cannot. 
If the number of generators $\Laxy$
is more than one, detecting a jump of $x_t$ does not tell us which 
generator  made $\psi_t$ jump. The jump of $x_t$ leaves the jump of 
$\psi_t$  in eq. \eqref{unrjumps} undetermined. This ambiguity can be fixed
in a natural class of special hybrid systems.

Let us add a vectorial structure $\{x^\al\}$ to the classical discrete
space and assume the following specific form of the hybrid generators $\Laxy$:
\beq\label{Lxnyn}
\Loa(x^\al,y^\al)\prod_{\be\neq\al}\delta(x^\be,y^\be),
\eeq
also with $\Loa(x^\al,x^\al)=0$.
The jump rates $T_\al(x^\al,y^\be)$ vanish if $\al\neq\be$, cf. eq. \eqref{rate}. 
Now, if we observe a jump of $x_t^\al$ we can uniquely determine the
jump of $\psi_t$. With the vectorized classical variables $x^\al$, 
the unraveling (\ref{unrdx}-\ref{unrjumps}) of the HME \eqref{HME} becomes unique
and the quantum trajectory can be monitored.

\section{From discrete to diffusive hybrid master equation}\label{sec_diffHME} 
Although discrete classical systems are important,   
most classical systems of interest,  like the Hamiltonian ones, 
are continuous. Therefore we are going to construct the
continuous limit of the obtained discrete HME \eqref{HME}. 
As is known, the only continuous classical Markovian process is
diffusion with additional deterministic drift. 
We start with the generic HME \eqref{HME} on the discrete
subset $\{x^n\}$ of the multidimensional continuum and generate 
a diffusion process in the continuous limit $\eps\rightarrow0$.

Note that some hybrid generators  can be of the form 
$\Layx=\hat{I}L_\al(x,y)\equiv L_\al(x,y)$, where $\hat I$
is the unit operator (and we used to spare them in notations). 
For some of such classical generators we introduce new label $n$
and the classical generators $L_n(x,y)$
So we have hybrid generators $\Laxy$ and the classical ones
$L_n(x,y)$.  The classical terms
are intended to generate diffusion terms 
for the classical subsystem in the continuous limit. The two classes 
of generators will be chosen as follows: %
\begin{widetext}
\bea
\Lo_\al(x,y)&=&
\Loa(y,y)\prod_n\frac{\delta(x^n-y^n,\eps)+\delta(y^n-x^n,\eps)}{\sqrt{2}},\label{La}
\label{Laeps}\\
L_n(x,y)&=&\frac{\delta(x^n-y^n,\eps)-\delta(y^n-x^n,\eps)}{\sqrt{2}\eps}
  \prod_{m\neq n}\frac{\delta(x^m-y^m,\eps)+\delta(y^m-x^m,\eps)}{\sqrt{2}}.\label{Ln}
\label{Lneps}
\eea
\end{widetext}
We introduce the positive semi-definite complex decoherence matrix $\DQ^{\al\be}$,
the positive semi-definite real diffusion matrix $\DC^{nm}$ and
the arbitrary complex matrix $\GCQ^{n\al}$ of backaction. 
Consider the following Hermitian block-matrix and let it be positive semi-definite:
\beq\label{bD}
\bD=\begin{bmatrix}\DQ&\GCQ\dg\\
                                                                \GCQ&\DC
                          \end{bmatrix}\geq0.
\eeq
We can see that nonzero backaction will require both nonzero decoherence
and diffusion. In addition to the constraints $\DQ\geq0,\DC=\DC\dg\geq0$
that we always take for granted, there are further constraints on matrices 
$\DQ,\DC,\GCQ$, equivalent to \eqref{bD}, to be shown in sec. \ref{sec_minHME}.                            

The following HME generates CP maps 
(since it reduces to the HME \eqref{HME} if we diagonalize $\bD$):
\begin{widetext}
\bea\label{prediffHME}
\frac{d\ro(x)}{dt}=-i[\Ho(x),\ro(x)]
&&+\DQ^{\be\al}\sum_y
\Bigl(\Laxy\ro(y)\Lo_\be\dg(x,y)-\Herm\Lo_\be\dg(x,y)\Layx\ro(x)\Bigr)\nn\\
&&+\DC^{nm}\sum_y
\Bigl(L_n(x,y)L_m(x,y)\ro(y)-L_n(y,x)L_m(y,x)\ro(x)\Bigr)\nn\\
&&+\GCQb^{n\al}\sum_y
\left(L_n(x,y)\Laxy\ro(y)-L_n(y,x)\Herm\Laxy\ro(x)\right).
\eea
\end{widetext}
In the continuous limit $\eps\rightarrow0$, the terms with $\Loa,\Lo_\be$ 
contribute to standard GKLS structures:
\beq
\Loa(x)\ro(x)\Lo_\be\dg(x)-\Herm\Lo_\be\dg(x)\Loa(x)\ro(x),
\eeq
with notation $\Loa(x)\equiv\Loa(x,x)$. %
The terms with $L_n,L_m$ yield diffusion of the classical variable $x$. 
With the notation $\eps_n$ meaning a vector whose only nonzero 
component is the $n'th$ one, and it is $\eps$, the yield at  $n\neq m$ reads:
\bea
&&\!\!\!\!\!\!\!\frac{
\ro(x\!+\!\eps_n\!\!+\!\eps_m)\!+\!\ro(x\!-\!\eps_n\!\!\!-\!\eps_m)
\!-\!\ro(x\!-\!\eps_n\!\!+\!\eps_m)\!-\!\ro(x\!+\!\eps_n\!\!\!-\!\eps_m)}{2\eps^2}\nn\\
&&\underset{\eps\rightarrow0}{\longrightarrow}
\frac12\der_n\der_m\ro(x),
\eea
where $\der_n\equiv\der/\der x^n$. %
We get the similar yield for $n=m$:
\beq
\frac{1}{2\eps^2}\Bigl(\ro(x+\eps_n)+\ro(x-\eps_n)-2\ro(x)
\Bigr)
\underset{\eps\rightarrow0}{\longrightarrow}
\frac12\der_n\der_n\ro(x).
\eeq
The cross-terms with $\Loa,L_n$ generate the non-trivial backaction of the quantum
system on the classical part: 
\bea
&\frac{1}{2\eps}\Bigl(\Loa(x+\eps_n)\ro(x+\eps_n)-\Loa(x-\eps_n)\ro(x-\eps_n)\Bigr)+\HC\nn\\
&\underset{\eps\rightarrow0}{\longrightarrow}
\der_n\left(\Loa(x)\ro(x)\right)+\HC
\eea
Using these limits in eq. \eqref{prediffHME} 
and  adding a deliberate classical drift of velocity $V(x)$,
we obtain the continuous limit of the discrete HME \eqref{HME}:
\begin{widetext}
\bea\label{diffHME}
\frac{d\ro(x)}{dt}=&&-i[\Ho(x),\ro(x)]
+\DQ^{\be\al}\Bigl(\Loa(x)\ro(x)\Lo_\be\dg(x)-\Herm\Lo_\be\dg(x)\Loa(x)\ro(x)\Bigr)\nn\\
&&+\tfrac12\DC^{nm}\der_n\der_m\ro(x)
+\left(\GCQb^{n\al}\der_n\left(\Loa(x)\ro(x)\right)+\HC\right)
-\der_n\left(V^n(x)\ro(x)\right).
\eea
\end{widetext}
The three \emph{constant}
parameter matrices  $\DQ,\DC,\GCQ$ are constrained by the
semi-definiteness $\bD\geq0$ of the block-matrix \eqref{bD} formed by them.

The above HME is not yet the general diffusive one. Obviously we can add 
an extra $x$-dependent decoherence $\Delta\DQ(x)$
as well as an extra diffusion $\Delta\DC(x)$
as long as $\bD\geq$ holds true after the replacements 
$\DQ\rightarrow\DQ+\Delta\DQ(x)$ and $\DC\rightarrow\DC+\Delta\DC(x)$.
That is, the validity of the diffusive HME \eqref{diffHME} extends for 
$x$-dependent parameters $\DQ(x)$ and $\DC(x)$ provided $\bD(x)\geq0$.
In the next section we show that also the backaction matrix $\GCQ$
can depend on $x$. 
 
\section{Covariant hybrid master equation}\label{sec_covHME}
The form \eqref{diffHME} of the HME is explicitly covariant under the
global linear transformations of the operator basis $\Loa(x)$ and the
classical variables $x$. We are interested in explicit covariant form under local,
i.e.: $x$-dependent complex linear transformation of the operator basis and
under general coordinate transformations of $x$. The form of such  
HME \eqref{diffHME} should be this (cf. \cite{oppenheim2022two}):
\begin{widetext}
\beq\label{covHME}
\frac{d\ro}{dt}=
-i[\Ho,\ro]+ \DQ^{\be\al}\Bigl(\Loa\ro\Lo_\be\dg-\Herm\Lo_\be\dg\Loa\ro\Bigr)
+\tfrac12\der_n\der_m(\DC^{nm}\ro)
+\der_n(\GCQb^{n\al}\Loa\ro+\HC)-\der_n\left(V^n\ro\right)
\eeq
\end{widetext}
where every object is function of $x$, the fact our above notations hide for the
sake of compactness.

\begin{small}
\emph{Correction.} 
The covariance of the Eq. (36) and its unravellings in Sec. IV.B
does not extend for  the general transformations of the classical variables but
for the linear ones.  To achieve general covariant forms, the  scalar hybrid density and 
the co(ntra)variant drift vector  should be used via two respective replacements:
\bea
\ro&\rightarrow&\frac{1}{\sqrtD}\ro,\nn\\
V^n&\rightarrow&V^n+\frac12\der_m\left(\sqrtD\DC^{nm}\right),\nn
\eea
where $g=1/\mathrm{det}\DC$. 
These replacements allow us to recast the non-covariant Fokker--Planck structure in Eq. (36) into 
the equivalent covariant one:
\bea
&&\frac{d\ro}{dt}=
-i[\Ho,\ro]+ \DQ^{\be\al}\Bigl(\Loa\ro\Lo_\be\dg-\Herm\Lo_\be\dg\Loa\ro\Bigr)\nn\\
&&+\frac{1}{2\sqrtD}\der_n\!\!\left(\sqrtD\DC^{nm}\der_m \ro\right)
\!+\!\frac{1}{\sqrtD}\der_n\!\!\left(\sqrtD(\GCQb^{n\al}\Loa\ro\!+\!\HC\!\!-\!\!V^n)\ro\right),\nn
\eea
where we recognize the first and second order covariant derivations in a Riemann
space of contravariant metric tensor $\DC^{nm}$.
For general covariance of the stochastic unraveling (49,50), the two Ito type SDEs should be changed 
for the equivalent Stratonovich ones and the additional constraint $\der_n(dW^n/\sqrtD)=0$ 
will have to be imposed \cite{diosi2023covariant}.  The general covariant formalism can
then be implemented in the rest of Sec. IV.B, the scope requires a separate work though.  
\end{small}

The coefficients $\DQ(x),\DC(x),\GCQ(x)$ satisfy the same constraint
\eqref{bD} $\bD(x)\geq0$ as before, now understood for all $x$: 
\beq\label{bDx}
\bD(x)=\left[\!\begin{array}{c|c}
\\
\DQ^{\al\be}(x)&~~~\GCQb^{m\be}(x)~~~~\\
\\
\hline
\\
\\
\!\!\GCQ^{n\al}(x)\!\!&\DC^{nm}(x)\\
\\
\\
\end{array}\right]\geq0.
\eeq

We prove the equivalence of the covariant HME \eqref{covHME} 
with \eqref{diffHME}.
By suitable choice of the operator basis $\Lo_\al(x)$ and 
the classical coordinates $x^n$, one can always transform $\GCQ(x)$ into
constant matrix. This results in the  HME \eqref{diffHME} which,
as we argued there, is valid for $x$-dependent $\DQ,\DC$.

The covariant diffusive HME \eqref{covHME} is the most general continuous
HME to generate CP dynamics. 
Every object in it is $x$-dependent.
The condition (\ref{bDx}) is necessary and sufficient when the generators $\Loa(x)$
are linearly independent and also linearly independent of the unit operator
$\hat I$. %
The hybrid Hamiltonian $\Ho$ and the classical drift $V$ are arbitrary. 
The Hermitian matrix $\DQ\geq0$ of decoherence, the real matrix $\DC\geq0$ of 
diffusion must form the positive semi-definite block matrix \eqref{bDx} in corners with
the matrix $\GCQ$ (and $\GCQ\dg$) of backaction.
We list three useful
alternatives that can always be achieved by trasformation of the reference
frames: a fixed operator basis $\{\Loa\}$,
or simultaneously diagonal $\DQ$ and $\DC$ with zeros and ones,   
or $\GCQ$ with zeros and ones in the main diagonal and zeros elsewhere. 
(These coordinate transformations may request the embeddig of $x$
in higher dimensions then they are of.) 

In addition to the explicit covariance, there is a further `gauge' freedom,
the descendant of the shifts \eqref{shift} in the discrete HME: 
\bea\label{diffshift}
\Loa&\rightarrow&\Loa+\ell_\al,\nn\\
\Ho&\rightarrow&\Ho-i\frac{i}{2}\bigl(\ell_\al^\star\Loa-\HC\bigr),\nn\\
V^n&\rightarrow&V^\al-(\GCQb^{n\al}\ell_\al+\HC),
\eea
where $\ell_\al(x)$ is an arbitrary complex function. 

 \subsection{Minimum-noise threshold}\label{sec_minHME}
We are going to breakdown the condition $\bD\geq0$ \eqref{bDx} into constraints
between $\bD$'s building blocks. Necessary conditions are the following: 
\bea\label{ranges}
\range\DQ&\geq&\range\GCQ\dg\GCQ,\\
\range\DC&\geq&\range\GCQ\GCQ\dg.
\eea
They express that the range of decoherence $\DQ$ cannot be narrower
than the range of $\Loa$'s that are coupled to the classical $x^n$'s
by backaction $\GCQ$. And similarly, the range of classical diffusion $\DC$ 
cannot be smaller than the range of $x^n$'s that are coupled to the $\Loa$'s.
Nonzero backaction means mandatory noise: both decoherence and diffusion.

Suppose that we have a given nonzero matrix $\GCQ$ of backaction
and we are interested in a certain minimum of the total irreversibility,
i.e., a certain minimum of the block-matrix $\bD$, implying
a certain minimum of the decoherence $\DQ$ and diffusion $\DC$
as we see below. Obviously, the strict positivity $\bD>0$ means
more noise than the minimum. We are interested in 
the maximum degenerate $\bD$ meaning the lowest
$\rank\bD$ which is limited by the rank  $\rCQ=\rank\GCQ$, 
otherwise $\bD\geq0$ 
cannot be true. Therefore we define the minimum-noise threshold by this:
\beq\label{condmin}
\rank\bD=\rank\GCQ.
\eeq 
Then the inequalities \eqref{ranges} saturate, 
$\range\DQ=\range\GCQ\dg\GCQ\label{rDQ}$ and
$\range\DC=\range\GCQ\GCQ\dg$, also meaning that 
$\rank\DQ=\rank\DC=\rCQ$. Both the number of coupled independent 
generators $\Loa$ and coordinates $x^n$ coincide with $\rCQ$. 

For a given $x$, transform $\GCQ$ into the frame where its elements are zero 
expect for an $\rCQ\times\rCQ$ unit matrix in the upper-left corner.  
Then, because of 
\eqref{ranges}, both $\DQ$ and $\DC$ respectively must have an 
$\rCQ\times\rCQ$ strictly positive matrix in their upper-left corners
and zeros elsewhere. Recall the general condition $\bD\geq0$.
If we drop rows and columns of zeros then the said non-zero 
$\rCQ\times\rCQ$ sub-matrices, denoted invariably, must be
each other's inverses \cite{diosi1995twoplanck}:
\beq\label{DCDQunit}
\DC(x)\DQ(x)=I.
\eeq  
This is the condition of the minimum-noise threshold \eqref{condmin} 
in the special frame fitted to the backaction matrix. 

We can now identify quantitatively and directly
the lower bound of irreversibility that hybrid systems must undergo even 
if the quantum and classical subsystems were reversible in themselves. 
At a given backaction strength $\GCQ$ (scaled now to be the unity) and
at the minimum noise threshold, the quantum decoherence strength 
$\DQ$ and classical diffusion strength $\DC$ are inverses of each other,
lower decoherence requests higher diffusion and vice versa. 

It is possible to decompose the constraint $\bD\geq0$ as well as the
minimum noise condition \eqref{condmin} into covariant relationships
(i.e.: valid in any reference frame) for the three parameter matrices.
Two equivalent forms of $\bD\geq0$ are the following \cite{oppenheim2022two}:
\bea\label{Opp1}
\GCQ\frac{1}{\DQ}\GCQ\dg&\leq&\DC,\label{Opp1leqDC}\\
\GCQ\frac{1}{\DQ}\DQ\GCQ\dg&=&\GCQ\GCQ\dg;\label{Opp1rangeDQ}
\eea
and 
\bea\label{Opp2}
\GCQ\dg\frac{1}{\DC}\GCQ&\leq&\DQ,\label{Opp2leqDQ}\\
\GCQ\dg\frac{1}{\DC}\DC\GCQ&=&\GCQ\dg\GCQ,\label{Opp2rangeDC}
\eea
where $1/\DQ$ and $1/\DC$ are generalized inverses.  
The threshold condition $\rank\bD=\rank\GCQ$ of minimum noise
corresponds to the saturation of inequalities into equalities.
Note that 
$(1/\DQ)\DQ=\range\DQ$ and $(1/\DC)\DC=\range\DC$; hence
the second lines in eqs. (\ref{Opp1},\ref{Opp2}) correspond to the 
mentioned identities $\range\DQ=\range\GCQ\dg\GCQ$,
and $\range\DC=\range\GCQ\GCQ\dg\label{rDC}$, respectively.
Under the above covariant parametric constraints the covariant HME \eqref{covHME} 
and its unravellings (\ref{covunrdx},\ref{covunrdpsi})
include all possible dynamics that contain noise on and above the threshold of consistency. (The appendix \ref{app1} shows the sharpness of the above conditions
on a simplest HME.)

\subsection{Stochastic unravellings}\label{sec_covunr}
When we construct unravellings of the HME \eqref{covHME},
we follow the steps of sec. \ref{sec_unr} and construct the two correlated
stochastic processes for $x_t$ and $\psi_t$ satisfying the condition
of unraveling \eqref{unrcond}. The two processes will diffusive ones this time.
First, take the trace of the diffusive HME \eqref{covHME} and obtain the
classical Fokker--Planck equation of the classical subsystem:
\bea\label{FPE}
\frac{d\rho(x)}{dt}=\!\!\!\!\!\!&&\frac12\der_n\der_m\left(\DC^{nm}(x)\rho(x)\right)-\nn\\
&&\!\!\!\!\!-\der_n\! \left(V^n(x)\rho(x)-2\Real\GCQb^{n\al}(x)\langle\Loa(x)\rangle
\rho(x)\right),
\eea
where $\Real$ denotes the real part of the subsequent expression.
The unraveling of this equation will be a  unique Brownian motion with a unique 
$\psi$-dependent drift.
To unravel the quantum subsystem, we need the
non-Hermitian (frictional) hybrid Hamiltonian defined by
\begin{widetext}
\beq\label{diffHfr}
-i\Hfr (x)=-\frac12\DQ^{\be\al}\Bigl(\left(\Lo_\be\dg(x)-\langle\Lo_\be\dg(x)\rangle\right)
                                                                      \left(\Loa(x)-\langle\Loa(x)\rangle\right)
+\left(\langle\Lo_\be\dg(x)\rangle\Loa(x)-\HC\right)\Bigr).
\eeq
\end{widetext}
The hybrid unraveling consists of two correlated 
diffusive stochastic processes, one for $x_t$, another for $\psi_t$:
\beq
dx^n=V^n(x)dt-2\Real\GCQb^{n\al}(x)\langle\Loa(x)\rangle dt-dW^n(x)
,\label{covunrdx}
\eeq
\vskip-16pt
\beq
d\psi\!=\!-i(\!\Ho(\!x\!)\!+\!\Hfr (\!x\!))\psi dt\! +\!(\Loa(\!x\!)\!-\!\langle\Loa(\!x\!)\rangle)\psi d\overline{\xi}^\al(\!x\!),\label{covunrdpsi}   
\eeq
where $dW^n(x)$ is real, and $d\xi^\al(x)$ is complex zero-mean Ito-differential of 
auxiliary stochastic processes, correlated as follows:  
\bea\label{dWdxi}
dW^n dW^m&=&\DC^{nm}dt,\nn\\
d\xi^\al d\overline{\xi}^\be&=&\DQ^{\al\be}dt,\\
dW^n d\xi^\al&=&\GCQ^{n\al}dt.\nn
\eea
Let us use the vector symbols $dW,d\xi$, then we get the equivalent
compact form of correlations:
\beq\label{dWdxibD}
 \begin{bmatrix}d\xi d\xi\dg&d\xi dW^T\\dW  d\xi\dg&dW dW^T\\ 
 \end{bmatrix}
=\bD dt.
\eeq

We prove that the unraveling (\ref{covunrdx},\ref{covunrdpsi})
satisfies the condition \eqref{unrcond}.
Like in sec. \ref{sec_unr}, we use the notation $\Po=\psi\psi\dg$ and
the same form \eqref{unrcondP} of the condition to be proved:
\beq
d\ro(z,t)=d \Mean\Po_t\delta(z-x_t).
\eeq
The Ito-differential on the rhs contains three terms and we are going to
express them by the equations of $d\psi$ and $dx$ of the unraveling
(see also \cite{layton2022healthier}). 
First, to calculate $d\Po=d\psi \psi\dg+\psi d\psi\dg+d\psi d\psi\dg$ 
we use the stochastic equation \eqref{covunrdpsi} of $d\psi$:
\bea\label{dP}
d\Po=&&-i[\Ho(z),\Po]dt\nn\\
&&+D_Q^{\be\al}\left(\Loa(x)\Po\Lo_\be\dg(x)-\Herm\Lo_\be\dg(x)\Loa(x)\Po\right)dt\nn\\
&&+\Bigl(\bigl(\Loa(x)-\langle\Loa(x)\rangle\bigr)\Po d\overline{\xi}^\al(x)+\HC\Bigr)dt.
\eea
Second, we calculate $d\delta(z-x)$ using stochastic equation \eqref{covunrdx} of $dx$:
\begin{widetext}
\beq\label{dd}
d\delta(z-x)
=\tfrac12\DC^{nm}\der_n\der_m\delta(z-x)dt
+\left(V^n(x)-2\Real\GCQb^{n\al}\langle\Loa(x)\rangle\right)
\der_n\delta(z-x)dt+\left(\der_n\delta(z-x)\right)dW^n(x),
\eeq
\end{widetext}
where the partial derivations refer to $x$ obviously. 
From here, we get the three terms of $d\Mean\Po\delta(z-x)$ after 
using $\der\delta(z-x)/\der x^n=-\der\delta(z-x)/\der z^n$, 
then taking the stochastic  mean over $dW,d\xi$ and $x$:
\begin{widetext}
\bea\label{dMean}
\Mean d\Po\delta(z-x)&=&-i[\Ho(z),\ro(z)]dt
+\DQ^{\be\al}(z)\left(\Loa(z)\ro(z)\Lo_\be\dg(z)-\Herm\Lo_\be\dg(z)\Loa(z)\ro(z)\right)dt,
\nn\\
\Mean \Po d\delta(z-x)
&=&\tfrac12\der_n\der_m(\DC^{nm}(z)\ro(z))
-\der_n\bigl(V^n(z)
-2\Real\GCQb^{n\al}(z)\langle\Lo_\al(z)\rangle\ro(z)\Bigr)dt,\\
\Mean d\Po d\delta(z-x)
&=&\der_n\Bigl(\GCQb^{n\al}(z)
\left(\Loa(z)-\langle\Loa(z)\rangle\right)\ro(z)\Bigr)dt
                                     +\HC\nn
\eea
\end{widetext}
By taking the sum of these three equations  we recognize that $\Mean\Po\delta(z-x)$
satisfies the HME \eqref{covHME}.

The hybrid unravelling corresponds to the time-continuous measurement
of the observables $\left(\GCQb^{n\al}\Lo_\al(x)+\HC\right)$. 
Hybrid unravelling contains an autonomous drift of the classical variables 
and general feedbacks:
the Hamiltonian, the decoherence matrix, the measured observable, and the measurement noise can depend on the measured signal $x$. 
These dependences could be part of time-continuous measurement but usually they are not.   
(Except for typical feedback Hamiltonians, linear in $dx/dt$.)   

As we can easily inspect, the unraveling (\ref{covunrdx},\ref{covunrdpsi}) 
is invariant for the shifts \eqref{diffshift}. 
But in general, it is not covariant under the linear transformations of the
operator basis $\{\Loa(x)\}$ of the HME! Therefore the unraveling of the
HME is not unique, inherits the ambiguity of unravellings \cite{wiseman2001diosi} 
of the quantum mechanical GKLS dynamics.
Observe that we left  the non-Hermitian correlation $d\xi^\al(x)d\xi^\be(x)$ unspecified whereas the stochastic processes $(x_t,\psi_t)$ depend on it. 
At deliberate choices of $d\xi^\al d\xi^\be$ we get different 
unravellings of the same HME \eqref{covHME}.  

Covariance of the unravelling can be achieved if the complex noise $d\xi^\al$
is covariant. The simplest way is  if we set 
\beq
d\xi^\al d\xi^\be=0,
\eeq
 see \cite{gisin1992quantum,percival1998quantum}, also \cite{diosi1986}.
Another option of covariant (and real) $d\xi^\al$ will be shown in sec. \ref{sec_mondiffHME}.
 
Just like unravellings of standard GKLS master equations, the hybrid unravellings
are not necessarily be in terms of pure states $\psi_t$.
We shall discuss generic mixed state unravellings later in this section. 
Before that, a specific case is considered which is a closest generalization 
of the pure state unravelings (\ref{covunrdx},\ref{covunrdpsi}).
Extension
from pure states $\psi_t$ for mixed states $\sio_t$ is straightforward since
the formalism is very similar. The pure state 
density operator $\Po$ gives way to $\sio$ and the equation of $d\psi$
is replaced  by an equation of $d\sio$. Accordingly, the definition
of unraveling reads
\beq
\ro(z,t)=\Mean\sio_t\delta(z-x_t),
\eeq
The eq. \eqref{covunrdx} of $dx^n$ is the same as before, 
the eq. \eqref{covunrdpsi} give way to the equation of $d\sio$:
\begin{widetext}
\beq\label{unrdsio}
d\sio=-i[\Ho(x),\sio]dt+\DQ^{\be\al}
\left(\Loa(x)\sio\Lo_\be\dg(x)-\Herm\Lo_\be\dg(x)\Loa(x)\sio\right)dt 
+\left((\Loa(x)-\langle\Loa(x)\rangle)\sio d\overline{\xi}^\al(x)+\HC\right)dt.   
\eeq
\end{widetext}
With the replacement $\Po\rightarrow\sio$, 
the equations of (\ref{FPE}-\ref{dMean}) of the previous proof apply exactly
in the same form and conclude that the process $(x_t,\sio_t)$ unravels the HME \eqref{covHME}. Notice the purification feature of \eqref{unrdsio} known from standard unravellings of GKLS master equations. If the state is not pure, 
i.e.: $\tr\sio^2<1$ then  
\begin{widetext}
\bea
\frac{d}{dt}\Mean\tr\sio^2&=&\frac{2}{dt}\Mean\tr(\sio d\sio)+\frac{1}{dt}\tr (d\sio)^2=\nn\\
&=&2\DQ^{\be\al}\tr\left(
\sio\Loa\sio\Lo_\be\dg-\Herm\sio\Lo_\be\dg\Loa\sio\right)
+\tr\left((\Loa-\langle\Loa\rangle)\sio^2(\Lo_\be\dg-\langle\Lo_\be\dg\rangle)\right)
\nn\\
&=&2\DQ^{\be\al}\tr\left(
\left(\sqrt{\sio}(\Lo_\be-\langle\Lo_\be\rangle)\sqrt{\sio}\right)\dg
\left(\sqrt{\sio}(\Loa-\langle\Loa\rangle)\sqrt{\sio}\right)
\right)>0.
\eea
\end{widetext}
An arbitrary initial mixed state $\sio_t$ will be purified asymptotically until $\tr\sio^2=1$ and then the mixed state eq. \eqref{unrdsio} becomes
equivalent with  the pure state eq. \eqref{unrdpsi}.
 
Given a covariant HME \eqref{covHME}, the family of unravellings
is larger than the above family of perfect purifying ones.
There are partially purifying unravellings if the HME is above the 
threshold of minimum noise, cf. \eqref{Opp2}:
\beq
\DQ=\GCQ\dg\frac{1}{\DC}\GCQ+\Delta\DQ\equiv D_\mathrm{Qmin}+\Delta\DQ
\eeq
where $\Delta\DQ\geq0$. Then the eq. \eqref{unrdsio} of mixed state
unravelling remains the same but the correlation $d\xi d\xi\dg=\DQ dt$
will be reduced to the minimum noise threshold 
\beq\label{dxidximin}
d\xi d\xi\dg=D_\mathrm{Qmin}dt=\GCQ\dg\frac{1}{\DC}\GCQ dt.
\eeq
The other correlations $dWd\xi=\GCQ dt$ and $dWdW^T=\DC dt$ are unchanged. 
The price we pay for the reduced noise is illustrated if we group the terms
of \eqref{unrdsio} as follows: 
\begin{widetext}
\bea\label{mixunrdsio}
d\sio=-i[\Ho(x),\sio]dt
&+&\tfrac12 D_\mathrm{Qmin}^{\be\al}\left(\Loa(x)\sio\Lo_\be\dg(x)-\Herm\Lo_\be\dg(x)\Loa(x)\sio\right)dt 
+\left((\Loa(x)-\langle\Loa(x)\rangle)\sio d\overline{\xi}^\al(x)+\HC\right)dt\nn\\   
&+&\tfrac12\Delta\DQ^{\be\al}\left(\Loa(x)\sio\Lo_\be\dg(x)-\Herm\Lo_\be\dg(x)\Loa(x)\sio\right)dt .
\eea
\end{widetext}
The first line corresponds to the perfect purifying unravelling, at the minimum
of noise \eqref{dxidximin} whereas the decoherence term in the second line counters
the purification. A highly mixed state becomes purer, a low level of mixture
becomes higher.  Mixedness may have a stationary value for certain HMEs
and certain unravellings.  We notice that the family of mixed state unravellings
is even larger since we can always set
\beq
d\xi d\xi\dg=\eta D_\mathrm{Qmin},~~~~(0 < \eta \leq 1).
\eeq

\subsection{Monitoring the diffusive quantum trajectory}\label{sec_mondiffHME}
We are going to show that, similarly to the jump trajectories in
sec. \ref{sec_mon}, also diffusive quantum trajectories $\psi_t$ 
can be monitored if the classical trajectories $x_t$ are observed.
As we shall see, this option of monitoring constrains the parameters 
of the HME \eqref{covHME} and singles out a unique unraveling
among the infinite many.
 
Monitoring  the   quantum trajectory $\psi_t$
via monitoring the classical $x_t$ is possible if and only if $d\psi$ \eqref{covunrdpsi}
uniquely depends on $dx$ \eqref{covunrdx}. Hence, the vector $d\xi$ must be 
a linear function of the vector $dW$: 
\beq\label{dxiFQCdW}
d\xi=\FQC dW.
\eeq
This deterministic relationship  removes the ambiguity of the unravelling
because it also specifies $d\xi d\xi$ that was free correlation,
see eqs. \eqref{dWdxibD}, in general unravelings. 
The above relationship should be consistent with the  correlations 
\eqref{dWdxibD}.
They  imply two equations: $\DQ=\FQC\DC\FQC\dg$ and $\GCQ\dg=\FQC\DC$. 
These equations possesses the solution:
\beq\label{FQC}
\FQC=\GCQ\dg\frac{1}{\DC},  
\eeq
and a constraint on the HME's parameters:
\beq\label{condmon}
\DQ=\GCQ\dg\frac{1}{\DC}\GCQ.
\eeq
This constraint is the condition that monitoring $\psi_t$ be possible.
Namely, we insert the solution \eqref{FQC} into \eqref{dxiFQCdW}
to obtain the desired map of $dW$ into $d\xi$: 
\beq\label{dxifromdW}
d\xi=\GCQ\dg\frac{1}{\DC}dW.
\eeq
It is remarkable that this equation defines covariant  $d\xi$, removes
the ambiguity of $d\xi d\xi$.

The condition \eqref{condmon} of monitoring coincides with the 
saturated condition \eqref{Opp2leqDQ} but, importantly,
the other condition \eqref{Opp2rangeDC} of minimum noise
is not necessary for monitoring. Accordingly, the option of
monitoring is still guaranteed above the noise threshold: $\range\DC$ can be
larger than $\range\GCQ\GCQ\dg$. At some $x$'s, there can be some
componenents of $x$ that are not coupled to the quantum subsystem,
represent above threshold noise, redundant for and do not prevent 
the monitoring of $\psi_t$. 

Using \eqref{dxifromdW}, we can eliminate $d\xi$ from the equations
of unraveling. Then both $dx$ and $d\psi$ (or $d\sio$) are driven
by the same real noise $dW$. Accordingly, the eq. \eqref{covunrdpsi}
becomes the following \cite{layton2022healthier}:
\begin{widetext}
\beq\label{moncovunrdpsi}
d\psi=-i\left(\Ho(x)+\Hfr (x)\right)\psi dt +dW^n(x)[\DC^{-1}(x)]_{nm}\GCQ^{m\al}(x)
\left(\Loa(x)-\langle\Loa(x)\rangle\right)\psi.
\eeq
\end{widetext}
In the special reference frame where
$\GCQ^{n\al}=\delta^{n\al}$ and we 
consider the minimum noise threshold where the non-degenerate
$\DQ,\DC$ are real and inverses of each other, we have  
$d\xi=\DC^{-1}dW=\DQ dW$, the eqs. 
(\ref{covunrdx},\ref{covunrdpsi}) 
of unraveling reduce to the following ones :
\bea
dx^\al&=&V^\al(x)dt-2\Real\langle\Loa(x)\rangle dt +dW_\al(x),
\label{mondx}\\
d\psi&=&-i(\Ho(x)-i\Hfr (x))\psi dt\nn\\ 
&&+\bigl(\Loa(x)-\langle\Loa(x)\rangle\bigr)\psi \DQ^{\al\be}dW_\be(x)\nn\label{mondpsi}.   
\eea 
Notice the new notation $dW_\al\equiv dW^n\vert_{n=\al}$.
We recognize the equations of correlated time-continuous
measurements (monitoring) of the observables $\Loa+\HC$
where $x_t$ is the measured signal. Here the model is a bit more
general because the signal can have an autonomous drift,  the 
Hamiltonian and the monitored observables can  
depend on the measured signal.

\section{Discussion}\label{sec_disc}
In this work we revisited, clarified and completed earlier results
on the Markovian master and stochastic equations of hybrid 
quantum-classical dynamics, paying attention to simplicity and brievity. 

The starting concept and the derivation of the HME in sec. \ref{sec_HME}
is most similar to the rigorous formulations of Blanchard and Jadczyk 
\cite{blanchard1993,blanchard1995}. The shift-invariance \eqref{shift} of the
HME  and the related nonuniqueness of the unraveling was not 
recognized before. Using vectorized classical variables is useful alternative
to the sophisticated conditions of  monitoring the jump quantum trajectory
in \cite{oppenheim2023objective}. 

Derivation in sec. \ref{sec_diffHME}  of the diffusive HME \eqref{diffHME} from the discrete \eqref{HME}
is completely new, in  attempt to replace the complicated though perhaps more
rigorous procedure of Oppenheim et al. \cite{oppenheim2022two}.  
Our derivation is based on the discrete forerunners of $\delta(x-y)$ and $\der/\der x^n$.
These are nontrivial while allowing for an elementary 
derivation. Importantly, the naive choice $\Laxy=\Loa(x)\delta(x-y)$ and 
$L_n(x,y)=\ketx\der_n\brax$ instead of (\ref{Laeps},\ref{Lneps}) 
turn out to be incorrect because off-block-diagonal terms play a role 
\footnote{J. Oppenheim private communication.}. 
The naive choice gave correct structure of the HME 
but below the correct threshold of minimum noise by a factor $1/2$. 

Sec. \ref{sec_covHME} re-derives the HME \eqref{covHME}
which was derived already by Oppenheim et al. 
\cite{oppenheim2022two} and Layton et al. \cite{layton2022healthier}. 
These works did not mention the covariance of their result, neither
put it in the usual form of co- and contravariant indices. 
Our work emphasizes and exploits that the HME is explicit 
covariant for local linear (i.e.: not necessarily unitary) transformations of the
Lindblad generators and for general transformations of the classical
coordinates. 

Sec. \ref{sec_minHME} presents  a pretty compact condition \eqref{condmin}
of the threshold for minimum noise. 
The phenomenon and equation \eqref{DCDQunit} of trade-off between
decoherence and diffusion was recognized in \cite{diosi1995twoplanck}, 
and has been extended recently for for the general diffusive HME in 
\cite{oppenheim2022two}, see eqs. (\ref{Opp1leqDC}-\ref{Opp2rangeDC})
using general matrix inverses to cover degenerate matrices of 
decoherence, diffusion, and backaction; their degeneracies are not exceptions
but typical. 

Our important new contribution in sec. \ref{sec_covunr}
is that the pure state diffusive unravellings of a diffusive HME are always possible and 
are exactly as ambiguous as the standard unravellings of GKLS master equations.
Since the ambiguity coincides
with that of the unravelings of pure quantum GKLS master equations, 
we can fix them in the same way. 
The choice $d\xi d\xi=0$ is well-known in theory of quantum state 
diffusion, used for covariance in \cite{diosi1986} and
developed by Gisin and Percival \cite{gisin1992quantum,percival1998quantum}
in the GKLS and Ito fomalisms. The full multitude of unravellings, applicable to
the hybrid dynamics as well, is discussed by Wiseman and the author
in \cite{wiseman2001diosi}.   

Sec. \ref{sec_mondiffHME} postulates the covariant condition \eqref{dxiFQCdW} 
of quantum trajectory monitoring, not used before. 
The resulting eqs. of monitoring coincide with those in \cite{layton2022healthier}.
Their claim that these eqs. are in one-to-one correspondence with the HME
is confirmed by covariance when the HME is at the threshold of minimum noise.  
The option of monitoring is not restricted for the HME of minimum noise, 
diffusion (not decoherence) can be higher than the threshold. 
  
\section{Conclusion}\label{sec_concl}
Rarely stated explicitly, but the  interaction between quantum and classical
systems has no other consistent mathematical model than time-continuous
quantum measurement and feedback, 
where measurement outcomes form the variables of
the classical system. This echoes von Neumann's visionary postulate.
To obtain a classical variable correlated with an unknown quantum state
the only consistent mathematical model is the von Neumann quantum measurement. Not too  surprizing, the equations of hybrid 
stochastic unravelings, both discrete  and continuous, 
coincide with the respective equations of time-continuous measurement, 
provided the classical system is identified with the measurement outcomes,
as in the elementary case \eqref{backaction}.
  
The unravellings (statistical interpretation) of hybrid master equations
are mathematical equivalents of time-continuous quantum measurements
as mentioned e.g. in \cite{diosi2014hybrid}.  
The advantage of hybrid master equations and unravellings over
time-continuous quantum measurement is not yet conceptual. 
The hybrid formalism may be fairly convenient in many 
applications, let them be e.g. foundations or improved semi-classical gravity.
No doubt, it may develop its own metaphysics as well. 

\begin{acknowledgments}
The author is obliged for the extended valuable discussions with Jonathan Oppenheim
and Isaac Layton.
This research was funded by the Foundational Questions Institute and Fetzer Franklin
Fund, a donor-advised fund of the Silicon Valley Community Foundation (Grant No's. FQXi-RFPCPW-2008,  FQXi-MGA-2103), the National Research, Development and Innovation Office (Hungary)
``Frontline'' Research Excellence Program (Grant No. KKP133827),
and the John Templeton Foundation (Grant 62099).
\end{acknowledgments}

\appendix
\section{Diffusive HME of two-level quantum system}\label{app1}
Consider a two-level quantum system coupled to a classical system of
a single variable $x$.
In Pauli's formalism, 
we can write the hybrid density \eqref{rox}  in the general form:
\beq\label{tlsrox}
\ro(x)=\tfrac12(1+\so(x))\rho(x),
\eeq
where $\so(x)=s_1(x)\sio_1+s_2\sio_2+s_3\sio_3$ and the length $s=\vert\sv\vert$
of the Bloch-vector $\sv=(s_1,s_2,s_3)$ must satify $s\leq1$.
Let the diffusive HME be the following simple one:
\beq\label{difftlsHME}
\frac{d\ro(x)}{dt}=\sio_3\ro(x)\sio_3-\ro(x)+G[\sio_3,\ro'(x)]_+ +\tfrac12\ro^{\prime\prime}(x).
\eeq
This corresponds to $\DQ=\DC=1$, and  $G$ is real. 
We prove that $\vert G\vert$ cannot be larger than $1$. 

Substitute eq. \eqref{tlsrox} and multiply both sides by $2$, yielding:
\begin{widetext}
\beq
\frac{d}{dt}\left((1+\so)\rho\right)
=[\sio_3(1+\so)\sio_3-(1+\so)]\rho+G[\sio_3,[(1+\so)\rho]']_++\tfrac12[(1+\so)\rho]^{\prr}.
\eeq
\end{widetext}
An equivalent form reads:
\beq
\frac{d\so}{dt}\rho+\so\frac{d\rho}{dt}
=-2\so_\perp\rho+2G\sio_3\rho'+\tfrac12\so\rho^{\prr}+\so'\rho'+\tfrac12\so^{\prr}\rho.
\eeq
We take trace of both sides, yielding the equation
$d\rho/dt=+2G(s_3\rho)'+\tfrac12\rho^{\prr}$. If we substitute it back, we get
\beq
\frac{d\so}{dt}\rho
=-2\so_\perp\rho+2G\sio_3\rho'+\so'\rho'+\tfrac12\so^{\prr}\rho-2G\so(s_3\rho)',
\eeq
where $\so_\perp=s_1\sio_1+s_2\sio_2$.
We multiply both sides by $\so$ and take $\tfrac12$ times their trace again:
\beq
\frac12\frac{ds^2}{dt}\rho
=-2s_\perp^2\rho+2Gs_3(1-s^2)\rho'+(s^2)'\rho'+\tfrac12\sv\sv^{\prr}\rho-2Gs^2s_3'\rho.
\eeq
Suppose $s^2=1$, then $ds^2/dt$ cannot be positive and $(s^2)'$ must vanish.
Thus we have the following inequality:
\beq
0\geq-2s_\perp^2+\tfrac12\sv\sv^{\prr}-2Gs^2s_3'.
\eeq
Now we insert the anzatz $\sv(x)=(\cos(x),0,\sin(x))$ into the inequality,
leading to
\beq
0\geq\frac12\frac{ds^2}{dt}=-2\cos^2(x)-2G\cos(x)-\tfrac12,
\eeq
which must be satisfied for all $x$.
This is equivalent with the upper bound on the backaction coupling:
\beq
G^2\leq1.
\eeq

\section{Hamiltonian HME}\label{app2}
Let $\Ho(x)$ be a our hybrid Hamiltonian where the classical subsystem
is canonical.
The first $N$ canonical variables $\{x^n; n=1,\dots,N\}$ are the coordinates
and the second $N$ ones $\{x^n; n=N+1,\dots,2N\}$ are the momenta.
The HME takes this form:
\beq\label{B1}
\frac{d\ro}{dt}=-i[\Ho,\ro]+\Herm\{\Ho,\ro\}+\dots~,
\eeq
where the ellipsis stands for mandatory decoherence and diffusion terms.
The bracket $\{\Ho,\ro\}$ is the Poisson bracket of classical canonical theory:
\beq\label{B2}
 \{\Ho,\ro\}=\varepsilon^{nm}(\der_n\Ho)(\der_m\ro),
\eeq
where $\varepsilon^{nm}$ is the $2N\times2N$ symplectic matrix.
To determine the decoherence and diffusion terms, we note  
the hybrid Hamiltonian can always have the following form:
\beq\label{B3}
\Ho(x)=\HoQ+\HCl(x)+\HoCQ(x)
\eeq
with
\beq\label{B4} 
\HoCQ(x)=h^\al(x)\Loa,
\eeq
where $h^\al(x)$'s are real functions and the $\Loa's$ are linearly independent 
Hermitian operators, also linearly independent of $\hat I$. 
Accordingly, the HME contains a purely Hamiltonian term $-i[\HoQ+\HoCQ,\ro]$,
a purely classical term $\{\HCl,\ro\}$,  
a backaction term $\Herm\{\HoCQ,\ro\}$, and  the mandatory decoherence and 
diffusive terms. It is easy to see that $\{\HCl,\ro\}$ corresponds to classical drift velocity 
$V^n=\varepsilon^{nm}\der_m\HCl$. By identifying the backaction term 
\beq\label{B5}
\Herm\{\HoCQ,\ro\}=\Herm\{h^\al\Loa,\ro\}
=\varepsilon^{nm}\Herm(\der_n h^\al)\Loa(\der_m\ro)
\eeq
in eq. (\ref{B1}) with the covariant HME's
backaction term $2\Herm\der_n(\GCQb^{n\al}\Loa\ro)$,
we read out the matrix of backaction:
\beq\label{B6}
\GCQ^{n\al}=\GCQb^{n\al}=-\tfrac12\varepsilon^{nm}\der_m h^\al.
\eeq
Hence, the Hamiltonian HME takes this general form:
\begin{widetext}
\beq\label{canHME}
\frac{d\ro}{dt}=
-i[\HoQ+\HoCQ,\ro]+\{\HCl,\ro\}+\Herm\{\HoCQ,\ro\}
+\DQ^{\be\al}\Bigl(\Loa\ro\Lo_\be\dg-\Herm\Lo_\be\dg\Loa\ro\Bigr)
+\tfrac12\der_n\der_m(\DC^{nm}\ro),
\eeq
\end{widetext}
and the matrices $\DQ,\DC,\GCQ$ should satisfy the non-negativity constraints
discussed in secs.~\ref{sec_diffHME},\ref{sec_covHME}. 
Note that the decoherence and diffusion
terms do not retain the covariance for canonical transformations
of the classical coordinates $\{ x^n\}$.
 
Let us apply this Hamiltonian HME  to the simple hybrid system of 
two identical harmonic oscillators, one is quantized the other is classical. 
The classical canonical variables are $(x^1,x^2)\equiv(q,p)$, 
the quantized ones are $(\Qo,\Po)$.  Let the hybrid Hamiltonian (\ref{B3}) consist of
\bea
\HoQ&=&\tfrac12(\Qo^2+\Po^2),\\
\HCl(q,p)&=&\tfrac12(q^2+p^2),\\
\HoCQ(q,p)&=&g(q\Qo+p\Po).
\eea
In expansion (\ref{B4}) of $\HoCQ$ we choose $\Lo_1=\Qo,~\Lo_2=\Po$ 
then $h^1=gq=gx^1,~h^2=gp=gx^2$ 
hence $G^{12}=-G^{21}=g/2$ while $G^{11}=G^{22}=0$.
The minimum noise condition reads, e.g., $G\DQ^{-1}G^T=\DC$ which means
that both $\DQ$ and $\DC$ are $2\times2$ diagonal matrices and satisfy 
\beq\label{B11}
\DQ^{11}\DC^{22}=\DQ^{22}\DC^{11}=g^2/4.
\eeq 
The resulting HME of $\ro(q,p)$ takes this form:
\begin{widetext}
\bea
\frac{d\ro}{dt}&=&
-i[\tfrac12(\Qo^2+\Po^2)+g(q\Qo+p\Po),\ro]
+(q\der_p\ro-p\der_q\ro+g\Herm(\Qo\der_p\ro-\Po\der_q\ro)\nonumber\\
&&+\tfrac12\DQ^{11}[\Qo,[\Qo,\ro]]+\tfrac12\DQ^{22}[\Po,[\Po,\ro]]
+\tfrac12\DC^{11}\der_q^2\ro+\tfrac12\DC^{22} \der_q^2\ro.
\eea
We see that both $\Qo$ and $\Po$ contribute to non-vanishing decoherence terms
and both $q$ and $p$ must diffuse. As we noticed, the mandatory 
decoherence and diffusion terms violate the canonical invariance of the
classical subsystem. And,  of course, they remain non-unique even
after taking the constraints (\ref{B11}) into the account .
\end{widetext}

\bibliography{diosi2023}{}
\end{document}